\documentclass[5p,sort&compress]{elsarticle}

\usepackage{amsmath}
\usepackage{amssymb}
\usepackage{bm}
\usepackage{graphicx}
\usepackage[pdftex,colorlinks]{hyperref}
\usepackage{mathrsfs}

\let\de=\partial
\DeclareMathOperator{\re}{Re}
\DeclareMathOperator{\tr}{tr}
\DeclareMathOperator{\Tr}{Tr}
\newcommand{\imag}{\text{i}}
\newcommand{\Da}{\mathscr{D}}
\newcommand{\Fa}{\mathscr{F}}
\newcommand{\La}{\mathscr{L}}
\newcommand{\Ha}{\mathscr{H}}
\newcommand{\dd}{\text{d}}
\newcommand{\he}[1]{#1^\dagger}
\newcommand{\vek}[1]{\bm{#1}}

\title{Chiral soliton lattice phase in warm QCD}
\author[1]{Tom\'a\v{s} Brauner\corref{cor}}
\ead{tomas.brauner@uis.no}
\author[1]{Helena Kole\v{s}ov\'{a}}
\ead{helena.kolesova@uis.no}
\author[2]{Naoki Yamamoto}
\ead{nyama@rk.phys.keio.ac.jp}
\cortext[cor]{Corresponding author}
\address[1]{Department of Mathematics and Physics, University of Stavanger, 4036 Stavanger, Norway}
\address[2]{Department of Physics, Keio University, Yokohama 223-8522, Japan}

\begin{document}

\begin{abstract}
We analyze the phase diagram of quantum chromodynamics at low-to-moderate temperature, baryon chemical potential and external magnetic field within chiral perturbation theory at next-to-leading order of the derivative expansion. Our main result is that the anomaly-induced chiral soliton lattice (CSL) phase is stabilized by thermal fluctuations. As a consequence, the CSL state may survive up to temperatures at which chiral symmetry is restored.
\end{abstract}


\maketitle


\section{Introduction}
\label{sec:intro}

Mapping the phase diagram of quantum chromodynamics (QCD) is an integral part of the quest to understand the fundamental laws of nature. The natural parameters used to specify the thermodynamic state of quark matter have traditionally been temperature $T$  and baryon chemical potential $\mu$, or density, with characteristic scales set by QCD dynamics. More recently, the magnetic field $B$ got into the spotlight~\cite{Andersen2016a,Miransky2015a} due to its relevance for the two natural laboratories used to study hot and dense quark matter: heavy-ion collisions and compact stars.

In spite of nearly a half century of effort, our understanding of the QCD phase diagram remains rudimentary. The problem is twofold. First, the strong dynamics of QCD pushes the validity of perturbative analytic approaches out to temperature and density regions beyond the reach of current experiments. Second, the notorious sign problem precludes lattice Monte Carlo simulations of QCD at moderate-to-high baryon densities. Consequently, much of what we know about the QCD phase diagram relies on qualitative insight and simplified models.

Based on previously published work~\cite{Son2008a}, two of us predicted in Ref.~\cite{Brauner2017a} a novel inhomogeneous phase of QCD, induced by the chiral anomaly in presence of a sufficiently strong magnetic field. The phase was dubbed chiral soliton lattice (CSL) in analogy with a similar phenomenon in magnetism~\cite{Togawa2012a,Kishine2015a}. In Refs.~\cite{Tatsumi2015a,Ferrer2017a,Ferrer2018a,Abuki2018a}, related anomaly-induced states of quark matter were studied in the limit of massless quarks within simplified models of QCD. Thanks to the universality of the chiral anomaly, CSL is in fact ubiquitous in gauge theories featuring confined fermionic degrees of freedom and external fields. The range of realizations of the CSL state now includes QCD matter under rotation without~\cite{Huang2018a} and with~\cite{Nishimura2020a} strangeness, dense matter under external magnetic field in an infinite class of QCD-like theories with (pseudo)real quarks~\cite{Brauner2019c,Brauner2019b}, and even the QCD vacuum in a driving laser field~\cite{Yamada2021a}. The first attempts to study the competition between CSL and nuclear matter were made in Refs.~\cite{Kawaguchi2018a,Chen2021a}.

Importantly, the prediction of the CSL phase is based on the low-energy effective field theory of QCD: the chiral perturbation theory (ChPT). It is therefore model-independent and can be considered a prediction of QCD itself, up to corrections that are calculable within a systematic derivative expansion. Previous work on CSL was however limited to the leading order (LO) of the derivative expansion, and thus to zero temperature.

In this letter, we extend the previous analysis to the next-to-leading order (NLO) by including one-loop corrections to the free energy from both quantum and thermal fluctuations. Our motivation is twofold. First, it is known that in three spatial dimensions, inhomogeneous phases with one-dimensional modulation suffer from the so-called Landau-Peierls instability due to thermal fluctuations of the order parameter. Hence for a number of candidate inhomogeneous phases in the QCD phase diagram~\cite{Buballa2015a}, only quasi-long-range order may be sustained~\cite{Baym1982a,Lee2015a,Hidaka2015b,Pisarski2019a}. On the other hand, it has been demonstrated that the Landau-Peierls instability can be avoided by explicitly breaking rotational invariance by an external magnetic field~\cite{Ferrer2020a}. For the same reason, it was argued previously~\cite{Brauner2017a,Brauner2017b} that this instability does not affect the CSL state. Yet, to the best of our knowledge, this is the first time to take a complete one-loop account of fluctuations of the order parameter in an inhomogeneous phase of QCD with a genuine long-range order.

The second piece of motivation behind this letter is the desire to map the domain in the phase diagram of QCD, occupied by the CSL phase. This is not merely an academic question; the infrared stability of the phase does not guarantee that it survives up to phenomenologically interesting temperatures. In this regard, our main result is that thermal fluctuations do not tend to melt, but instead further stabilize the CSL state. It is therefore conceivable that the CSL phase stretches all the way to the temperatures at which chiral symmetry is restored, although such a conjecture cannot be tested within ChPT.

The rest of the text is organized as follows. Section~\ref{sec:CSLreview} offers the reader a brief review of the origin of the CSL phase and its analysis at the classical, tree level, corresponding to the LO of the derivative expansion of ChPT. The basic setup for the calculation of free energy at one loop (NLO) is explained in Sec.~\ref{sec:setup}. The NLO analysis requires renormalization. While this is standard material, we provide some details in Sec.~\ref{sec:renormalization} for the reader's convenience. The main result of the paper is contained in Sec.~\ref{sec:FatNLO}. We focus here on elaborating its consequences. Further details together with a complementary analysis of the CSL phase will appear in a forthcoming publication. Finally, in Sec.~\ref{sec:discussion} we summarize and conclude.

Throughout the paper, we use the natural units $\hbar = c = 1$ and set the elementary electric charge to $e=1$.


\section{Brief review of the CSL phase}
\label{sec:CSLreview}

The low-energy dynamics of QCD is dominated by the Nambu-Goldstone bosons of the spontaneously broken chiral symmetry. At the lowest energies, only the two lightest quark flavors, and hence only the three pions, are relevant. The starting point of the tree-level derivation of the CSL phase is therefore the Lagrangian of two-flavor ChPT,
\begin{equation}
\La_\text{LO}=\frac{f_\pi^2}4\left[\tr(D_\mu\he\Sigma D^\mu\Sigma)+2m_\pi^2\re\tr\Sigma\right]+\La_\text{WZW}.
\label{LagLO}
\end{equation}
The sole parameters of ChPT at this order are the pion mass $m_\pi$ and decay constant $f_\pi$. The latter determines the intrinsic cutoff scale of ChPT as $4\pi f_\pi$. Furthermore, $\Sigma$ is the $2\times2$ unitary matrix variable of ChPT, the covariant derivatives of which include minimal coupling to external fields, if present. We are only going to need coupling to the electromagnetic field, for which $D_\mu\Sigma=\de_\mu\Sigma-\imag A_\mu[\tfrac{\tau_3}2,\Sigma]$, where $A_\mu$ is the electromagnetic gauge potential and $\tau_3$ the third Pauli matrix. Finally, $\La_\text{WZW}$ is the Wess-Zumino-Witten (WZW) term that incorporates the effects of the chiral anomaly~\cite{Wess1971a,Witten1983a}. In presence of the electromagnetic gauge field, $A_\mu$, and an auxiliary gauge field, $A_\mu^\text{B}=(\mu,\vek0)$, coupling to baryon number, it can be expressed as~\cite{Son2008a}
\begin{equation}
\La_\text{WZW}=\left(\frac{A_\mu}2-A_\mu^\text{B}\right)j_\text{GW}^\mu,
\label{WZWterm}
\end{equation}
where
\begin{align}
j^\mu_\text{GW}=-\frac1{24\pi^2}\epsilon^{\mu\nu\alpha\beta}\tr\biggl[&(\Sigma D_\nu\he\Sigma)(\Sigma D_\alpha\he\Sigma)(\Sigma D_\beta\he\Sigma)\\
\notag
&+\frac{3\imag}4F_{\nu\alpha}\tau_3(\Sigma D_\beta\he\Sigma+D_\beta\he\Sigma\Sigma)\biggr]
\end{align}
is the topological Goldstone-Wilczek current~\cite{Goldstone1981a}.

In magnetic fields $B\gtrsim m_\pi^2$, charged pions are expected to decouple due to Landau level quantization. Only the neutral pions then survive. Inserting accordingly $\Sigma=e^{\imag\tau_3\phi}$, with $\phi\equiv\pi^0/f_\pi$ being a dimensionless neutral pion field, in the Lagrangian~\eqref{LagLO} and assuming from now on a constant uniform background magnetic field $\vek B$, we obtain
\begin{equation}
\La_\text{LO}^{(\pi^0)}=\frac{f_\pi^2}2(\de_\mu\phi)^2+m_\pi^2f_\pi^2\cos\phi+\frac{\mu}{4\pi^2}\vek B\cdot\vek\nabla\phi.
\label{LagLOpi0}
\end{equation}
This was the starting point of the derivation of the CSL ground state in Ref.~\cite{Brauner2017a}. It is obvious without any calculation that in the chiral limit ($m_\pi\to0$), the last, anomalous term in Eq.~\eqref{LagLOpi0} induces a condensate with a constant gradient of $\phi$ in an arbitrarily weak magnetic field. This state was studied previously in Refs.~\cite{Bergman2009a,Thompson2008a,Rebhan2009a} and is sometimes referred to as the meson supercurrent.

A detailed analysis of the classical field theory defined by Eq.~\eqref{LagLOpi0} shows that away from the chiral limit, the trivial chiral-symmetry-breaking vacuum with $\phi=0$ (hereafter referred to as the normal phase) is only stable in magnetic fields below
\begin{equation}
B_\text{CSL}=\frac{16\pi m_\pi f_\pi^2}\mu.
\label{BCSL}
\end{equation}
For $B>B_\text{CSL}$ the CSL state is energetically favored over the normal phase. A detailed account of the properties of the CSL ground state can be found in Ref.~\cite{Brauner2017a}. Here we will focus on the limiting case of CSL that forms the ground state at the phase boundary, $B=B_\text{CSL}$, namely the domain wall,
\begin{equation}
\phi_0(z)=4\arctan e^{m_\pi z},
\label{wall}
\end{equation}
where we have chosen the orientation of the Cartesian coordinate system so that the magnetic field points along the positive $z$-semiaxis. The classical free energy per unit area of the domain wall relative to the normal phase is
\begin{equation}
\frac{\Fa_0}S=8m_\pi f_\pi^2-\frac{\mu B}{2\pi},
\label{FLO}
\end{equation}
whence the critical field~\eqref{BCSL} is easily recovered.

The domain wall~\eqref{wall} will play a key role throughout this paper. This is a topological defect whose appearance characterizes the phase transition from the normal phase to the CSL phase. Further increasing the magnetic field beyond $B_\text{CSL}$ results in the formation of a ``stack of parallel domain walls''~\cite{Son2008a,Eto2013a}, which is nothing but CSL. If we are only interested in the localization of the phase boundary separating the CSL phase from the normal phase, then the competition of the domain wall with the trivial QCD vacuum is all we need to worry about. This is the strategy that we employ below when we consider the effect of fluctuations. A more detailed analysis, including loop corrections to the CSL ground state inside the CSL phase, will appear in a follow-up paper.


\section{Setup of the calculation}
\label{sec:setup}


\subsection{Power-counting scheme}
\label{subsec:counting}

To make sure that the contributions of fluctuations to the free energy are under control, we need a power-counting scheme. We employ a modification of the standard power counting of ChPT~\cite{Scherer2012a} in which
\begin{equation}
\de_\mu,m_\pi,T,A_\mu=\mathcal O(p^1),\qquad
A_\mu^\text{B}=\mathcal O(p^{-1}).
\label{powercounting}
\end{equation}
In this power counting, the part of the WZW term~\eqref{WZWterm} proportional to $\mu$ is of order $\mathcal O(p^2)$, which makes the previous tree-level analysis of CSL~\cite{Brauner2017a} consistent. Note that the assignment of a negative counting order to $\mu$ is only possible thanks to the fact that the WZW term is the sole place in the ChPT Lagrangian to all orders where $\mu$ appears. One can also view this as indicating that baryon degrees of freedom are heavy and decouple from our effective theory. The baryon density carried by CSL, on the other hand, has a positive counting order and thus is naturally small compared to the cutoff scale of ChPT. Within the scheme~\eqref{powercounting}, one-loop corrections to free energy, generated by the $\mathcal O(p^2)$ Lagrangian, will be suppressed compared to tree-level free energy by a factor of $\varepsilon\sim[p/(4\pi f_\pi)]^2$, where $p$ is the characteristic scale of $m_\pi$, $T$ or $\sqrt{B}$.

In order to be able to pin down the phase boundary between the normal and CSL phases at NLO, we only need to calculate the difference of their one-loop free energies. Suppose that we were able to evaluate the free energy as a (generally nonlocal) functional of the neutral pion background $\phi$. Consider a systematic expansion of this functional in powers of the small expansion parameter $\varepsilon$,
\begin{equation}
\Fa[\phi]=\Fa_0[\phi]+\varepsilon^1\Fa_1[\phi]+\varepsilon^2\Fa_2[\phi]+\dotsb.
\end{equation}
We can expand similarly the configuration $\bar\phi$ of lowest free energy, that is the ground state,
\begin{equation}
\bar\phi=\phi_0+\varepsilon^1\phi_1+\varepsilon^2\phi_2+\dotsb.
\end{equation}
Then at the NLO of the expansion, the free energy of the ground state is given by
\begin{equation}
\Fa[\bar\phi]=\Fa_0[\phi_0]+\varepsilon^1\Fa_1[\phi_0]+\mathcal O(\varepsilon^2).
\label{Fgeneral}
\end{equation}
The naively expected contribution $\int\dd x\,\varepsilon^1\phi_1\frac{\delta\Fa_0[\phi_0]}{\delta\phi}$ is absent since the LO ground state $\phi_0$ is a stationary state of the functional $\Fa_0[\phi]$.

Equation~\eqref{Fgeneral} is analogous to the usual rule for the first-order perturbative correction to energy eigenvalues in quantum mechanics. Its validity thus relies on the assumption of a nondegenerate spectrum. Indeed, the domain wall~\eqref{wall} is a unique one-dimensional solution of the LO equation of motion for the Lagrangian~\eqref{LagLOpi0} carrying a unit quantum of topological charge, up to trivial degeneracy due to translation invariance. The important message is that at NLO of the derivative expansion of ChPT, we do not need to know the loop corrections to the domain wall profile. All we need to do is to evaluate the one-loop free energy of ChPT on the LO background~\eqref{wall}.


\subsection{Fluctuation determinant}
\label{subsec:fluctuation}

To find the one-loop free energy, we need to calculate the determinant of the differential operator that defines the part of the classical Lagrangian quadratic in fluctuations around the chosen stationary state. In our case, the one-loop free energy consists of contributions from the neutral and charged pion fluctuations above the domain wall background,
\begin{equation}
\beta\Fa_\text{1-loop}=\frac12\Tr\log\Da^{(\pi^0)}+\Tr\log\Da^{(\pi^\pm)},
\end{equation}
where $\beta\equiv1/T$ is the inverse temperature and $\Tr$ a trace in the operator sense. The neutral pion fluctuation operator $\Da^{(\pi^0)}$ follows at once from Eq.~\eqref{LagLOpi0},
\begin{equation}
\Da^{(\pi^0)}=\Box+m_\pi^2\cos\phi_0.
\label{Dpi0}
\end{equation}
The fluctuation operator in the charged pion sector can be deduced by expanding the full Lagrangian~\eqref{LagLO} to second order in charged pion fields, following the same steps as in Appendix C of Ref.~\cite{Brauner2017a}. Here we neglect the contribution of the WZW term to the bilinear Lagrangian for charged pions as it is of order $\mathcal O(p^4)$ in our power-counting scheme~\eqref{powercounting}. Choosing the vector potential for the external magnetic field as $\vek A=(0,Bx,0)$ then gives
\begin{equation}
\Da^{(\pi^\pm)}=\Box+2\imag Bx\de_y+B^2x^2-(\phi_0')^2+m_\pi^2\cos\phi_0,
\label{Dpipm}
\end{equation}
where the prime indicates a derivative with respect to $z$.

Separation of variables into longitudinal and transverse components with respect to the magnetic field vector $\vek B$ now leads, with a little effort, to the one-loop free energy
\begin{align}
\label{F1loop}
\beta\Fa_\text{1-loop}={}&\frac12\sum_{n,\vek p_\perp,\lambda}\log[\omega_n^2+\vek p_\perp^2+m_\pi^2(1+\lambda)]\\
\notag
&+\frac{BS}{2\pi}\sum_{n,m,\lambda}\log[\omega_n^2+(2m+1)B+m_\pi^2(1+\lambda)],
\end{align}
where the first and second term corresponds respectively to the contribution of neutral and charged pions, $n$ labels (bosonic) Matsubara frequencies, $\vek p_\perp$ transverse momentum of neutral pion excitations, and $m$ Landau levels of charged pion excitations. Finally, the index $\lambda$ runs over the spectrum of an effective Hamiltonian $\Ha$ describing motion in the $z$-direction on the domain wall background. With the explicit domain wall profile~\eqref{wall} at hand, one finds that this effective Hamiltonian is of the P\"oschl-Teller type,
\begin{equation}
\Ha_n=-\de_Z^2-\frac{n(n+1)}{\cosh^2Z},
\label{effHam}
\end{equation}
where $Z\equiv m_\pi z$ is a dimensionless coordinate and $n=1,2$ for neutral and charge pion excitations, respectively.

The sums over Matsubara frequencies and Landau levels as well as the integral over transverse momentum can be carried out, or at least simplified, using standard techniques. The bulk of the computation of the one-loop free energy on the domain wall background consists of an evaluation of the sum over $\lambda$. Using the well-known properties of the P\"oschl-Teller Hamiltonians~\eqref{effHam}, the sum can be carried out using the following master formulas:
\begin{equation}
\sideset{}{'}\sum_\lambda f(\lambda)=f(-1)-\frac2\pi\int_0^\infty\frac{f(P^2)}{1+P^2}\,\dd P
\label{lambdasumneutral}
\end{equation}
for $n=1$ and
\begin{align}
\label{lambdasumcharged}
\sideset{}{'}\sum_\lambda f(\lambda)={}&f(-4)+f(-1)\\
\notag
&-\frac1\pi\int_0^\infty\left(\frac2{1+P^2}+\frac4{4+P^2}\right)f(P^2)\,\dd P
\end{align}
for $n=2$. Both formulas are valid for smooth functions $f$ satisfying the convergence criterion $\lim\limits_{P\to\infty}f(P^2)/P=0$. Note the primes on the summation symbols, which indicate that the sums have been regularized by subtracting a sum over the spectrum of the ``free particle'' Hamiltonian $\Ha_0$. Using these master formulas together with Eq.~\eqref{F1loop} therefore yields directly the difference of free energies of the domain wall and the normal phase. Throughout the rest of the paper, we will usually implicitly have in mind this difference whenever we speak of the free energy (of the domain wall), without explicitly mentioning the subtraction of the normal phase contribution every time.

Before we can write down the final result for the one-loop free energy, we have to deal with the fact that its zero-temperature part is divergent and requires renormalization. While it would be possible to renormalize Eq.~\eqref{F1loop} ad hoc, it is a nontrivial consistency check to determine all the necessary counterterms separately. This is what we do in the next section.


\section{Vacuum renormalization}
\label{sec:renormalization}

At NLO of the derivative expansion, the two-flavor version of ChPT contains altogether 12 independent operators~\cite{Scherer2012a}. Fortunately, we do not need to carry out complete one-loop renormalization of ChPT. We only have to renormalize the one-loop effective action on a generic neutral pion background. Moreover, we are only interested in the difference of free energies of the domain wall and the normal phase. This reduces the required counterterms to
\begin{equation}
\begin{split}
\La_\text{c.t.}={}&\ell_1[(\de_\mu\phi)^2]^2+\ell_2(\de_\mu\phi)^2m_\pi^2\cos\phi\\
&+\ell_3m_\pi^4(\cos^2\phi-1),
\end{split}
\label{Lagct}
\end{equation}
where $\ell_{1,2,3}$ are dimensionless couplings to be fixed. These are related to the standard basis of NLO couplings through
\begin{equation}
\ell_1=l_1+l_2,\qquad
\ell_2=l_4,\qquad
\ell_3=l_3+l_4,
\label{Ltoell}
\end{equation}
following the notation of Ref.~\cite{Andersen2012b}. The divergent parts of $l_{1,2,3,4}$ can be fixed in the vacuum using the $\overline{\text{MS}}$ re\-nor\-ma\-li\-zation scheme with dimensional re\-gu\-la\-ri\-zation in $D\equiv4-2\epsilon$ spacetime dimensions. When the renormalization scale is chosen as $m_\pi$, they are given by~\cite{Andersen2012b}
\begin{equation}
l_i=-\frac{\gamma_i}{2(4\pi)^2}\left(\frac1\epsilon+1-\bar l_i\right),
\label{MSbar}
\end{equation}
where the algebraic coefficients $\gamma_i$ are~\cite{Gasser1984a}
\begin{equation}
\gamma_1=\frac13,\qquad
\gamma_2=\frac23,\qquad
\gamma_3=-\frac12,\qquad
\gamma_4=2.
\end{equation}
In the numerical results presented below, we use the following values of the finite counterterms,
\begin{equation}
\begin{aligned}
\bar l_1&=-0.4\pm 0.6,\qquad&
\bar l_2&=4.3 \pm 0.1,\\
\bar l_3&=3.53\pm 0.26,\qquad&
\bar l_4&=4.4\pm 0.2.
\end{aligned}
\label{Lbar}
\end{equation}
The values of $\bar l_{1,2,4}$ were determined phenomenolo\-gi\-cal\-ly~\cite{Colangelo2001a}, whereas the value of $\bar l_3$ is based on lattice simulations~\cite{Baron2010a,Aoki2020a}. We are only going to use the mean values of $\bar l_{1,2,3,4}$; the errors are displayed just for a rough indication of the uncertainty of our results.

Apart from the NLO operators that carry the divergences generated by one-loop diagrams with LO vertices, the operators in the LO Lagrangian themselves receive finite loop corrections. These are necessary in order to match the parameters $m_\pi,f_\pi$ to the physical values of the pion mass and decay constant, which we denote as $M_\pi,F_\pi$. The latter are identified by writing the part of the renormalized one-loop effective action of ChPT, quadratic in $\phi$, as $\frac{F_\pi^2}2(\de_\mu\phi)^2-\frac12M_\pi^2F_\pi^2\phi^2$, cf.~Eq.~\eqref{LagLOpi0}. A detailed calculation leads to the matching relations
\begin{align}
F_\pi^2&=f_\pi^2+\frac{m_\pi^2}{8\pi^2}\bar l_4,\\
M_\pi^2F_\pi^2&=m_\pi^2f_\pi^2-\frac{m_\pi^4}{32\pi^2}(\bar l_3-4\bar l_4).
\end{align}
Choosing $M_\pi=140\text{ MeV}$ and $F_\pi=92\text{ MeV}$ for the physical input, the mean values of the finite counterterms~\eqref{Lbar} then fix the parameters $m_\pi,f_\pi$ as
\begin{equation}
m_\pi\approx142\text{ MeV},\qquad
f_\pi\approx86\text{ MeV}.
\label{NLOparameters}
\end{equation}


\section{Free energy at one loop}
\label{sec:FatNLO}

It is convenient to split the calculation of the one-loop free energy using Eq.~\eqref{F1loop} into the zero-temperature and thermal parts. In the zero-temperature part, the integration over frequency $\omega$ and transverse momentum $\vek p_\perp$, as well as the sum over Landau levels, can be done in a closed form within dimensional regularization. What remains is the sum over $\lambda$, which is converted by Eqs.~\eqref{lambdasumneutral}--\eqref{lambdasumcharged} into an integral over the quasimomentum $P$. Here one has to perform suitable subtraction in order to extract the analytically calculable divergent part. Upon adding the contributions to the free energy from the counterterms~\eqref{Lagct}, the final result for the renormalized, zero-temperature free energy per unit area of the domain wall at NLO reads
\begin{align}
\notag
\frac{\Fa_1^{T=0}}S=&-\frac{m_\pi^3}{72\pi^2}(70-120\log2+16\bar l_1+32\bar l_2+3\bar l_3)\\
\notag
&+\frac{B^{3/2}}{\sqrt2\pi}\biggl\{\zeta(-\tfrac12,\tfrac12-\tfrac{3m_\pi^2}{2B})+\zeta(-\tfrac12,\tfrac12)\\
\notag
&-\int_0^\infty\frac{\dd P}{2\pi}\frac4{1+P^2}\biggl[\zeta\bigl(-\tfrac12,\tfrac12+\tfrac{m_\pi^2(1+P^2)}{2B}\bigr)\\
\label{FNLOT0}
&+\frac23\biggl(\frac{m_\pi^2}{2B}\biggr)^{3/2}(1+P^2)^{3/2}\biggr]\\
\notag
&-\int_0^\infty\frac{\dd P}{2\pi}\frac8{4+P^2}\biggl[\zeta\bigl(-\tfrac12,\tfrac12+\tfrac{m_\pi^2(1+P^2)}{2B}\bigr)\\
\notag
&+\left(\frac{m_\pi^2}{2B}\right)^{3/2}\biggl(\frac23(4+P^2)^{3/2}-3(4+P^2)^{1/2}\biggr)\biggr]\biggr\},
\end{align}
where $\zeta(s,q)\equiv\sum\limits_{m=0}^\infty\frac1{(m+q)^s}$ denotes the Hurwitz $\zeta$-function.

In the thermal part of the free energy, the Matsubara sum can be evaluated explicitly. The contribution of neutral pions can then, upon additional integration over transverse momentum, be brought to the form
\begin{align}
\label{FNLOTneutral}
\frac{\Fa_1^{T,(\pi^0)}}S=-\frac{\zeta(3)T^3}{2\pi}-\frac{m_\pi^2T}{\pi^2}&\int_0^\infty P\arctan P\\
\notag
&\times\log\Bigl(1-e^{-\beta m_\pi\sqrt{1+P^2}}\Bigr)\,\dd P.
\end{align}
In the contribution of charged pions, the sum over Landau levels and integral over $P$ remain unevaluated,
\begin{align}
\notag
\frac{\Fa_1^{T,(\pi^\pm)}}S=\frac{BT}\pi\sum_{m=0}^\infty\biggl\{&\log\left[1-e^{-\beta\sqrt{(2m+1)B-3m_\pi^2}}\right]\\
\label{FNLOTcharged}
&+\log\left[1-e^{-\beta\sqrt{(2m+1)B}}\right]\\
\notag
&-\frac1\pi\int_0^\infty\dd P\left(\frac2{1+P^2}+\frac4{4+P^2}\right)\\
\notag
&\qquad\qquad\times\log\left[1-e^{-\beta\epsilon(m,P^2)}\right]\biggr\},
\end{align}
where we have introduced the shorthand notation
\begin{equation}
\epsilon(m,\lambda)\equiv\sqrt{(2m+1)B+m_\pi^2(1+\lambda)}.
\end{equation}
The full NLO free energy of the domain wall equals the sum of the above three contributions,
\begin{equation}
\Fa_1=\Fa_1^{T=0}+\Fa_1^{T,(\pi^0)}+\Fa_1^{T,(\pi^\pm)},
\end{equation}
and constitutes the desired one-loop correction to Eq.~\eqref{FLO}.


\subsection{Asymptotic expansion of the free energy}

The free energy depends on three physical scales, $m_\pi$, $T$ and $\sqrt B$, and can by convention be expressed in terms of one of these and two dimensionless ratios. One may expect that the above-derived full result can be further simplified in case one or two of these scales is (are) much smaller than the other(s). Indeed, a simple analytical approximation is possible close to the chiral limit, that is when $m_\pi$ is much smaller than the other two scales.

Let us introduce the dimensionless ratio $\gamma\equiv\sqrt{B}/T$ along with the auxiliary function
\begin{align}
\varphi(\gamma)&\equiv\sum_{m=0}^\infty\int_0^\infty\frac{\dd t}{\omega_m(t)\bigl[e^{\gamma\omega_m(t)}-1\bigr]},\\
\omega_m(t)&\equiv\sqrt{2m+1+t^2}.
\end{align}
Upon some manipulation, the leading contribution to the free energy of the domain wall in the asymptotic expansion in $m_\pi \ll T,\sqrt{B}$ can be brought to the simple form,
\begin{equation}
\begin{split}
\frac{\Fa}S\simeq{}&8m_\pi f_\pi^2-\frac{\mu B}{2\pi}\\
&+\frac{3\log 2}{4\pi^2}m_\pi B-\frac{m_\pi T^2}6-\frac{6m_\pi B}{\pi^2}\varphi(\gamma).
\end{split}
\label{Fapprox}
\end{equation}
The three loop corrections on the second line arise, respectively, from zero-temperature charged pion fluctuations, thermal neutral pions, and thermal charged pions. Remarkably, all of these can be reproduced from the LO free energy on the first line of Eq.~\eqref{Fapprox} by taking into account finite $B$- and $T$-dependent corrections to the physical pion mass and decay constant~\cite{Agasian2001a}, 
\begin{align}
\notag
M_\pi^2(B,T)&=m_\pi^2\biggl[1-\frac{T^2}{24f_\pi^2}\\
&\qquad\quad-\frac{B\log2}{16\pi^2f_\pi^2}+\frac{B}{2\pi^2f_\pi^2}\varphi(\gamma)\biggr]+\dotsb,\\
F_\pi^2(B,T)&=f_\pi^2\left[1+\frac{B\log2}{8\pi^2f_\pi^2}-\frac{B}{\pi^2f_\pi^2}\varphi(\gamma)\right]+\dotsb.
\end{align}
This is a nontrivial consistency check of our result. Further analytical approximation is possible in the limits $\gamma\ll1$ and $\gamma\gg1$~\cite{Agasian2001a}. 


\subsection{Critical surface of the CSL phase}

We expect a phase transition between the normal phase and the CSL phase to occur on the critical surface in the $T$-$\mu$-$B$ space where the free energy of the domain wall relative to the normal phase vanishes. The baryon chemical potential only enters the free energy through the last, anomalous term in Eq.~\eqref{FLO}. Thus, the condition defining the critical surface is most easily expressed as an equation for the critical chemical potential as a function of $B,T$,
\begin{equation}\label{muCSL}
\mu_\text{CSL}=\frac{16\pi m_\pi f_\pi^2}B+\frac{2\pi}B\frac{\Fa_1}S.
\end{equation}
The NLO free energy $\Fa_1$ obviously defines, up to trivial rescaling, directly the shift of the critical surface in the $\mu$-direction due to quantum and thermal fluctuations.

\begin{figure}[t]
$$
\includegraphics[width=\columnwidth]{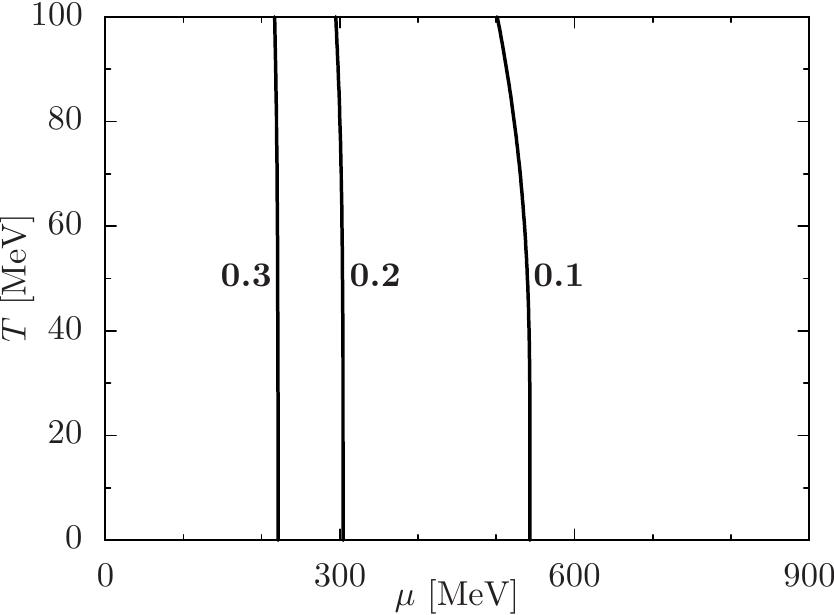}
$$
\caption{Critical curve separating the normal and CSL phases in the $T$-$\mu$ plane for several fixed values of $B$. The latter are given in $\text{GeV}^2$ and indicated in bold next to the corresponding curves.}
\label{fig:fixedB}
\end{figure}

Figures~\ref{fig:fixedB} and~\ref{fig:fixedT} display the critical curves as defined by Eq.~\eqref{muCSL} respectively in the $T$-$\mu$ plane for several fixed values of $B$, and in the $B$-$\mu$ plane for several fixed values of $T$. Some features of these numerical results may be understood based on the asymptotic expression~\eqref{Fapprox}. First, in strong magnetic fields ($B\gg m_\pi^2,T^2$), the NLO correction to free energy is dominated by zero-temperature charged pion fluctuations and is well approximated by the third term in Eq.~\eqref{Fapprox}. Hence, the phase transition shifts to higher chemical potentials compared to the LO result.

On the other hand, the negative terms on the second line of Eq.~\eqref{Fapprox} suggest (and the numerical results confirm) that \textit{thermal} corrections shift the phase transition towards smaller $\mu$, i.e.~they expand the domain in the phase diagram where the CSL phase lives. This is a consequence of different excitation spectra in the normal and CSL phases. In the normal phase, thermal excitations are exponentially suppressed at low temperatures due to the nonzero gap of both neutral and charged pions. The CSL phase, on the other hand, features a gapless excitation even away from the chiral limit: the phonon of the soliton lattice~\cite{Brauner2017a}.\footnote{In the limiting case of the domain wall, the phonon reduces to a bound state, localized on the domain wall in the $z$-direction and propagating only in the transverse directions. Its contribution to the thermal free energy gives the first term in Eq.~\eqref{FNLOTneutral}. This is the leading contribution to the thermal free energy when $T\ll m_\pi,\sqrt{B}$ (except for the limit $B\to 3m_\pi^2$ where also the lowest-lying charged pion excitation becomes light, see Sec.~\ref{sec:discussion}).} Hence at low temperatures, the thermal pressure of the CSL phase will dominate over that of the normal phase by mere Bose-Einstein statistics.


\section{Discussion and outlook}
\label{sec:discussion}

In this paper, we have analyzed the effect of quantum and thermal fluctuations on the phase transition between the normal (vacuum) and CSL phases in QCD in external magnetic field. Our calculation is based on the assumption that the phase transition occurs via the formation of a domain wall. This assumption raises several issues that require further investigation, some of which will be addressed in a forthcoming publication.

\begin{figure}[t]
$$
\includegraphics[width=\columnwidth]{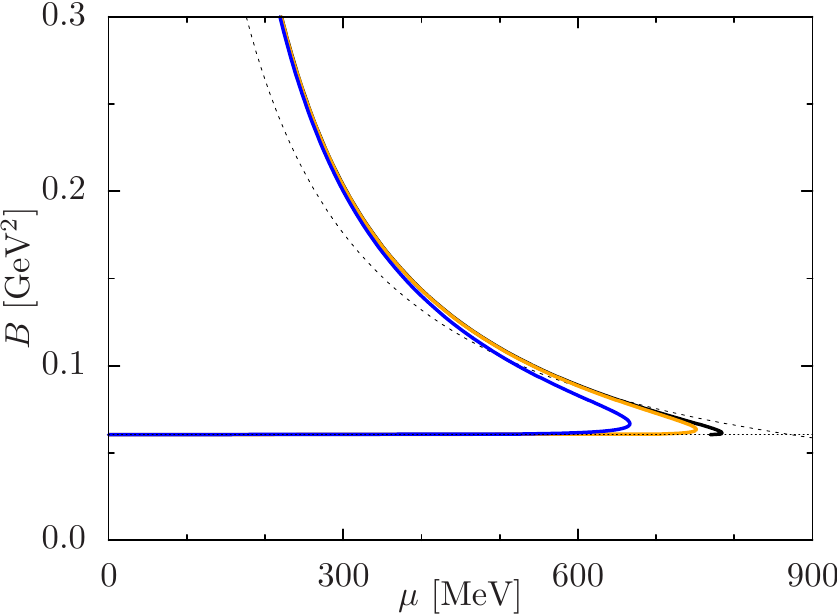}
$$
\caption{Critical curve separating the normal and CSL phases in the $B$-$\mu$ plane for several fixed values of $T$. The values of $T$ displayed are $0$ (thick black), $40$ (thick orange) and $80$ (thick blue) $\text{MeV}$. For comparison, the LO result~\eqref{BCSL} with parameters fixed by Eq.~\eqref{NLOparameters} is shown by the thin dashed line. The horizontal thin dotted line indicates the domain wall instability threshold, $B=3m_\pi^2$.}
\label{fig:fixedT}
\end{figure}

It is known that the domain wall~\eqref{wall} is unstable under charged pion fluctuations for $B<3m_\pi^2$. This has a direct bearing on our result for the domain wall free energy. The zero-temperature one-loop free energy~\eqref{FNLOT0} picks an imaginary part for $B<3m_\pi^2$ due to the presence of the $\zeta(-\frac12,\frac12-\frac{3m_\pi^2}{2B})$ term. Accordingly, the zero-temperature NLO critical curve, shown by the thick solid line in Fig.~\ref{fig:fixedT}, ends at the point $(\mu,B)=(769\text{ MeV},0.0605\text{ GeV}^2)$.

Even worse, the thermal free energy~\eqref{FNLOTcharged} of charged pion fluctuations of the domain wall, taken at face value, diverges as $B$ approaches $3m_\pi^2$ from above. This is an effect of the lowest-lying charged pion excitation, which is localized in all three dimensions and contributes the $m=0$ part of the first term in Eq.~\eqref{FNLOTcharged}.

It would be easy to discard the above features as mere signs of a breakdown of the derivative expansion of ChPT. Indeed, in a self-consistent treatment of the phase transition, the instability at $B<3m_\pi^2$ might in principle disappear if the fluctuation effects are strong enough to lift the lowest-lying charged pion bound state on the domain wall above certain threshold, or remove the bound state altogether. Alternatively, one might guess that the focus on the domain wall is misled and the fluctuation effects make the phase transition first-order, so that it proceeds directly from the normal phase to a CSL state with a finite lattice spacing. Such a fluctuation-induced first-order phase transition between homogeneous and inhomogeneous phases was discussed previously e.g.~in Refs.~\cite{Brazovskii1975a,Karasawa2016a,Yoshiike2017a}.

We believe, however, that the puzzling behavior of our results in weak magnetic fields reflects a lack of understanding of the interplay of CSL with charged pion degrees of freedom. It was observed already in Ref.~\cite{Brauner2017a} that there is an upper critical field at which the CSL itself becomes unstable under charged pion fluctuations. Our NLO result presented here indicates that a new phase with a mixture of neutral and charged pion condensates might occupy a larger region in the phase diagram than thought before.

In any case, the results of this letter confirm that in sufficiently strong magnetic fields, the chiral anomaly can turn the QCD vacuum into a periodic soliton. We actually find that thermal fluctuations further stabilize such inhomogeneous order. One might then speculate that the anomalous inhomogeneous phase survives up to temperatures at which chiral symmetry is restored. Thanks to the spontaneous breakdown of the exact translation invariance by such a phase, this intriguing possibility would imply the existence of a sharp phase transition at a temperature somewhere between $150$ and $200\text{ MeV}$, in contrast to the ordinary chiral crossover at zero magnetic field.


\section*{Acknowledgments}

We would like to thank Jens Oluf Andersen, Aleksi Kurkela and Andreas Schmitt for useful discussions and comments. This work was supported in part by the grant no.~PR-10614 within the ToppForsk-UiS program of the University of Stavanger and the University Fund, and by the Keio Institute of Pure and Applied Sciences (KiPAS) project at Keio University and JSPS KAKENHI Grant No.~19K03852.


\end{document}